\documentclass[longbibliography,slac_one]{revtex4}
\usepackage{graphicx}
\usepackage{fancyhdr}
\usepackage{subfigure}

\def \Bs {B_s^{0}}

\def \Bsb {\overline{B}{}_s^{0}}

\begin{document}

\title{$CP$ Violation in $\Bs \to J/\psi \phi$}

%

\author{Yuehong Xie}
\affiliation{University of Edinburgh, Edinburgh EH9 3JZ, UK}

\begin{abstract}
Study of $CP$ violation in the decay channel $\Bs \to J/\psi \phi$ is essential to exploring  and constraining physics beyond the Standard Model
in the quark flavour sector. The  experimental progress in this area of activity at the LHC and Tevatron is discussed.  
\end{abstract}

\maketitle

\thispagestyle{fancy}


\section{Probing new physics in $\Bs$--$\Bsb$ mixing  with  $\Bs \to J/\psi \phi$ }

The origins of $CP$ violation in fundamental physics theory remain a mystery.
The $\Bs$ system provides excellent laboratories to probe $CP$ violating new physics,
since new particles beyond the Standard Model (SM) may enter the loop-mediated  $\Bs$ meson mixing process, 
leading to discrepancies of $CP$ asymmetries  with their SM expectations. 

The effective Hamiltonian of  the $\Bs$--$\Bsb$ system can be written as
\begin{equation}
\mathbf{H_s} = \left( \begin{array}{cc}
M^s_{11} & M^s_{12}  \\
M^{s*}_{12} & M^s_{22}
\end{array} \right)
-
\frac{i}{2}\left( \begin{array}{cc}
\Gamma^s_{11} & \Gamma^s_{12}  \\
\Gamma^{s*}_{12} & \Gamma^s_{22}
\end{array}  \right) \,,
\end{equation}
where $M^s_{11}=M^s_{22}$ and $\Gamma^s_{11}=\Gamma^s_{22}$ hold under the assumption of $CPT$ invariance.
The off-diagonal  elements $M^s_{12}$ and $\Gamma^s_{12}$ are responsible for $\Bs$--$\Bsb$ mixing.
Diagonalizing the Hamiltonian matrix leads to the two mass eigenstates $B^s_{\rm H,L}$ (${\rm H}$ and ${\rm L}$ denote  heavy and light, respectively),
with mass $M^s_{\rm H,L}$  and decay width $\Gamma^s_{\rm H,L}$. $B^s_{\rm H,L}$  are  linear combinations of flavour eigenstates
 with    complex coefficients $p$ and $q$ that satisfy $\left| p \right|^2 + \left| q \right|^2 = 1$: $|B^s_{\rm L,H}\rangle = p |B_s\rangle \pm q |\bar{B}_s \rangle$.
New physics contribution in the mixing process could affect
the mass difference between the heavy and light mass eigenstates, $\Delta m_s \equiv M^s_{\rm H} - M^s_{\rm L} $,  
 the decay width difference between the light and heavy mass eigenstates, $\Delta \Gamma_s \equiv \Gamma^s_{\rm L} - \Gamma^s_{\rm H}$,
 and the semileptonic asymmetry $a_{SL}^s \approx \frac{|\Gamma^s_{12}|}{|M^s_{12}|}\sin\phi^s_{12}$, where
 $\phi^s_{12}\equiv \arg (-M^s_{12}/\Gamma^s_{12})$ is a convention-independent phase difference.

The decay $\Bs \to J/\psi \phi$ (charge conjugate is implied in this paper)
 proceeds dominantly via  a tree level $b \to c {\bar c } s$  diagram that is  well understood in the SM.
 This makes it an ideal place to search for demonstration of new physics  with only limited  hadronic uncertainties.
Ignoring the doubly Cabibbo-suppressed penguin contributions in the $b \to c {\bar c } s$ decay process, we denote the  phase difference
between the amplitude for a direct decay of $\Bs$ to a  $CP$ eigenstate $f$ with eigenvalue $\eta_f$
and the amplitude for decay after oscillation   as
\begin{equation}
 \phi_s \equiv -\arg \left( \eta_f \frac{q}{p}\frac{\bar A_f}{A_f} \right) \approx \arg\left(-M^s_{12}  \right) -2 \arg \left( V_{cb}V_{cs}^* \right) \,,
\end{equation}
where $A_f$ and $\bar A_f$ are the decay amplitudes of $\Bs \to f$  and $\Bsb \to f$, respectively.
(Discussions about controlling the effect of the penguin diagrams in the $b \to c {\bar c } s$ decay process
can be found in Section 3.2. of Ref.~\cite{Bediaga:2012py} and references therein.)
$\phi_s$ is very precisely predicted within the SM,  
$\phi_s^{\rm SM} = -2\beta_s =-2 \arg\left(-\frac{V_{ts}V_{tb}^*}{V_{cs}V_{cb}^*}   \right)  = -0.036 \pm 0.002$ rad~\cite{Lenz:2006hd, Charles:2011va}, 
however, it could be altered by new physics contribution in $\Bs$ mixing.
Note $\phi_s \approx \phi^s_{12}$ is a good approximation unless either the $b \to c {\bar c } s$ decay process
or $\Gamma^s_{12}$ is affected by physics beyond the SM. 

Neglecting the small $CP$ violation in $\Bs$ mixing, {\it i.e.} assuming $|q/p|=1$, we can write the mixing-induced $CP$ asymmetry  as $S_f = -\sin \phi_s$.
Thus $\phi_s$ can be extracted from the time-dependent $CP$ asymmetry in the decay $\Bs \to J/\psi \phi$, measurement of which requires to
identify the flavour of the initial $\Bs$ or $\Bsb$ mesons.
The final state of the decay $\Bs \to J/\psi \phi$
is an admixture of two $CP$-even  
and one $CP$-odd 
eigenstates.
There is  also a $CP$-odd final state due to  S-wave $K^+K^-$ contribution (specified using the subscript ``{\rm S}'') under the $\phi$ peak.
An angular analysis is needed to statistically disentangle the four $CP$ eigenstates~\cite{Dighe:1998vk, Dunietz:2000cr}. 
The differential  rates in decay time and angular variables  for the decay $\Bs \to J/\psi \phi$ are given in the references~\cite{Dighe:1998vk, Dunietz:2000cr, Xie:2009fs}. 
The time-dependent angular analysis of flavour tagged $\Bs \to J/\psi \phi$ decays  also needs to take into account 
the experimental effects such as background contamination, detector resolution and reconstruction efficiency.

\section{Historical review of the experimental study}

\begin{figure}[!htb]
\centering
\includegraphics[width=0.45\textwidth]{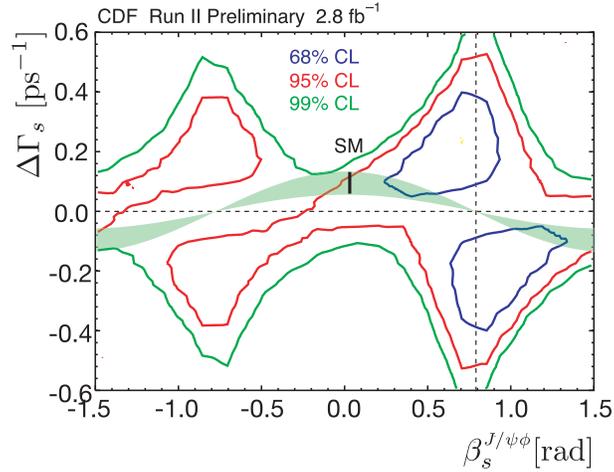}
\caption{
The confidence regions in the $\beta_s^{J/\psi\phi}$-$\Delta\Gamma_s$ plane in Ref.~\cite{D0-5928}
 from the combination of early D0 and CDF results each based on 2.8 fb$^{-1}$, where $\beta_s^{J/\psi\phi} = -\phi_s/2$. } 
\label{fig:tevatron_early}
\end{figure}

\begin{figure}
\centering
\includegraphics[width=0.45\textwidth]{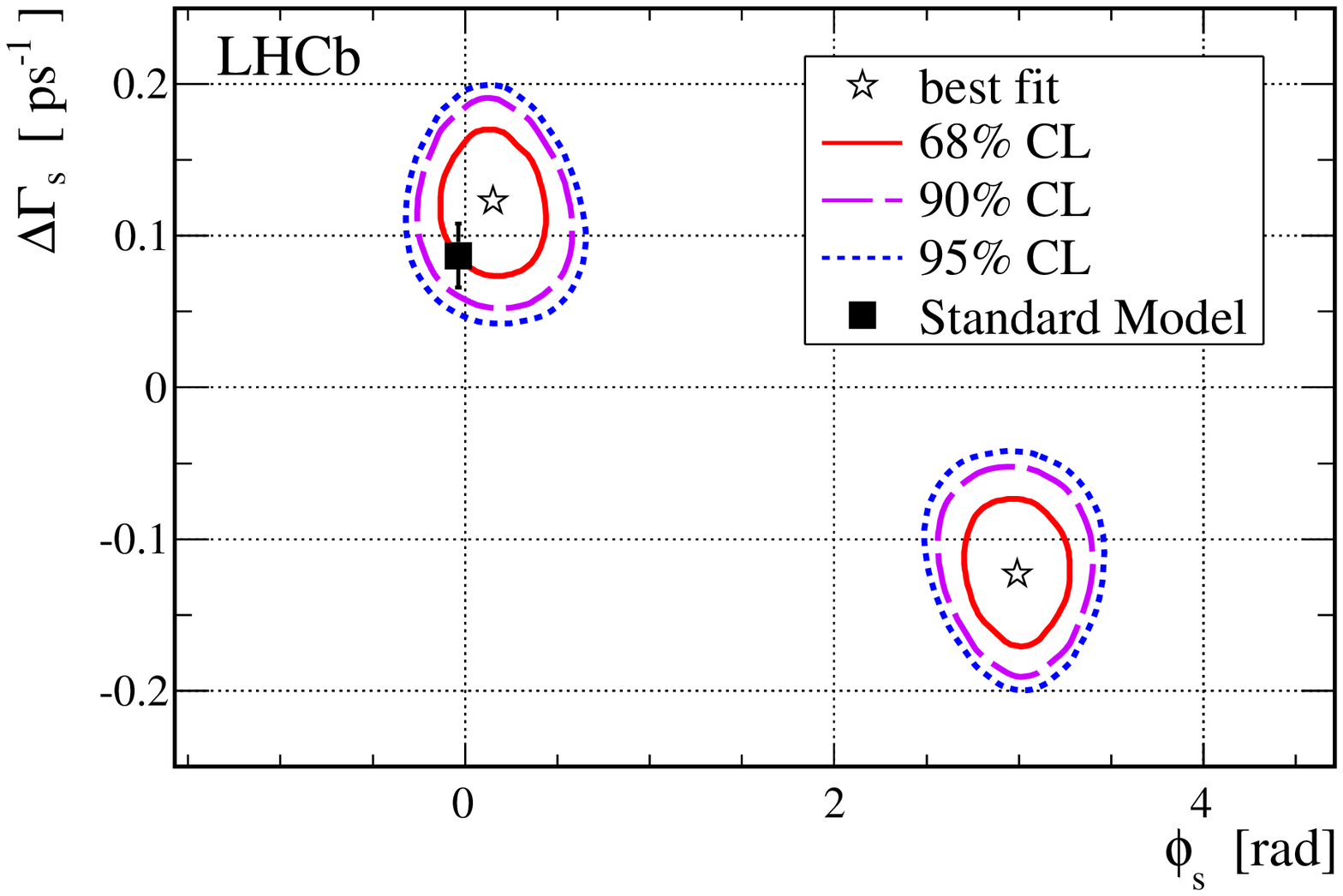}
\includegraphics[width=0.45\textwidth]{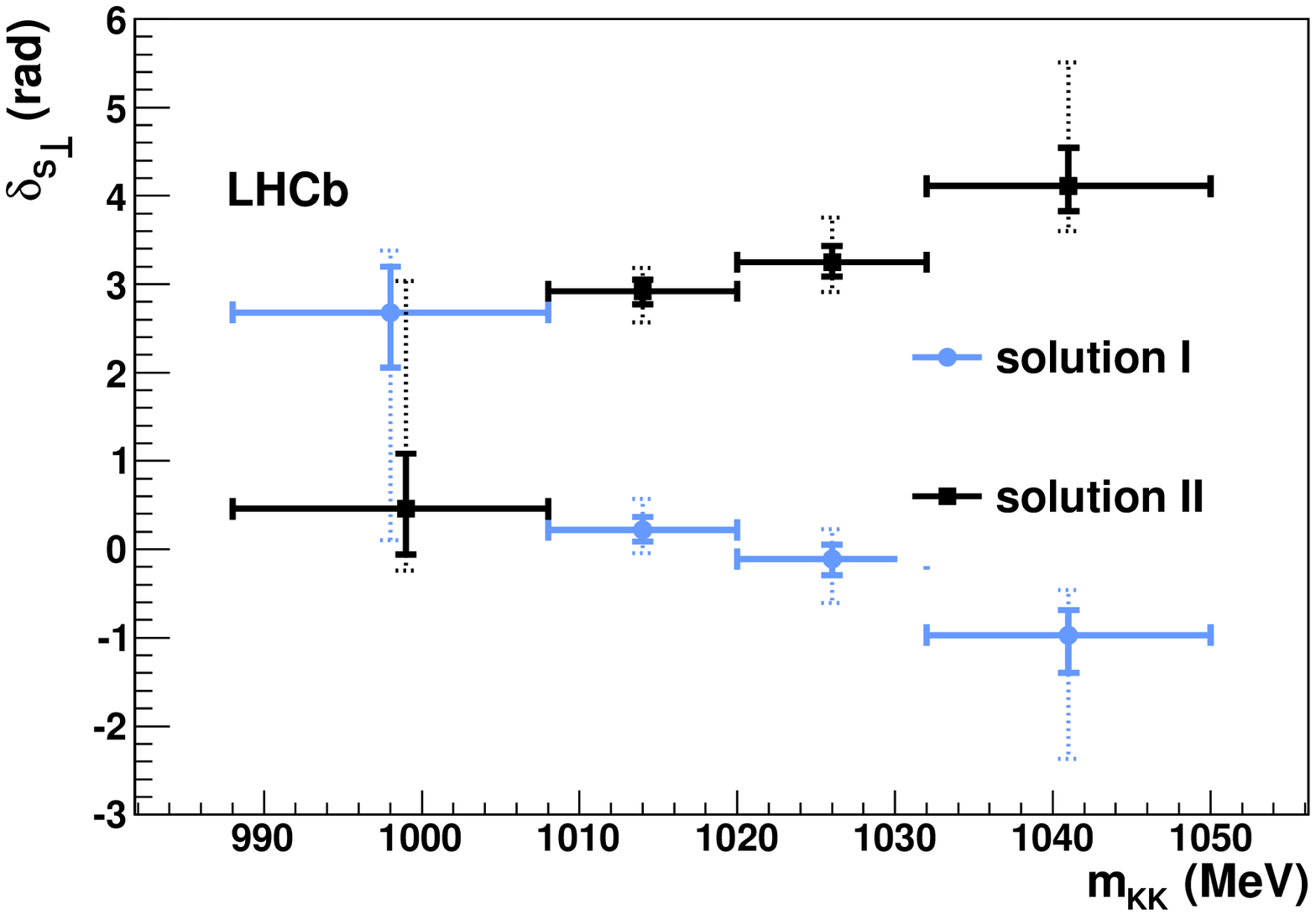}
\caption{
(Left) The confidence regions in the $\phi_s$-$\Delta\Gamma_s$ plane from the LHCb analysis of 0.37 fb$^{-1}$ ~\cite{LHCb-PAPER-2011-021}.
(Right) The phase difference between the S-wave and P-wave amplitudes as a function of the $K^+K^-$ invariant mass for each of the two 
 solutions in the $\Bs \to J/\psi \phi$ analysis~\cite{LHCb-PAPER-2011-028}. 
}
\label{fig:lhcb_early}
\end{figure}

The decay $\Bs \to J/\psi \phi$  has been extensively studied at the Tevatron and LHC experiments~\cite{D0-5928, CDF:2011af, Abazov:2011ry, LHCb-PAPER-2011-021}.
Early study at CDF and D0 experiments  each using an integrated luminosity of 2.8 fb$^{-1}$ showed a $2.1\sigma$ deviation from their SM 
expectations (Fig.~\ref{fig:tevatron_early})~\cite{D0-5928}. 
However, this was not confirmed by the CDF updated result using 5.2 fb$^-1$ of data~\cite{CDF:2011af} and the D0 updated result using 8.0 fb$^{-1}$ of data~\cite{Abazov:2011ry},
nor by the much more precise LHCb result based on 0.37 fb$^{-1}$ of $pp$ collision data (Fig. ~\ref{fig:lhcb_early} (left))~\cite{LHCb-PAPER-2011-021}.
Furthermore, following the method described in Ref.~\cite{Xie:2009fs},
LHCb  used the  0.37 fb$^{-1}$ sample   to measure  the phase difference between the S-wave and P-wave amplitudes as a function of the $K^+K^-$ invariant mass for each of the 
two ambiguous solutions (see Fig. ~\ref{fig:lhcb_early} (left))  and identified the solution with a decreasing trend as the physical solution 
(solution I in Fig.~\ref{fig:lhcb_early} (right)). 
This determined  the sign of $\Delta\Gamma_s$ to be positive with a 4.7 $\sigma$ significance~\cite{LHCb-PAPER-2011-028}, and resolved the two-fold
ambiguity of $\phi_s$ for the first time. The remaining solution of $\phi_s$ and $\Delta\Gamma_s$ in the LHCb analysis of $\Bs \to J/\psi \phi$  decays
was  consistent with the SM expectations. The most up-to-date results from the LHCb and CDF  experiments  will be discussed in  details in Section~\ref{sec:status}.

\section{Recent experimental progresses and implications}
\label{sec:status}

\begin{figure}[!htb]
\centering
\includegraphics[width=0.42\textwidth,height=5.5cm]{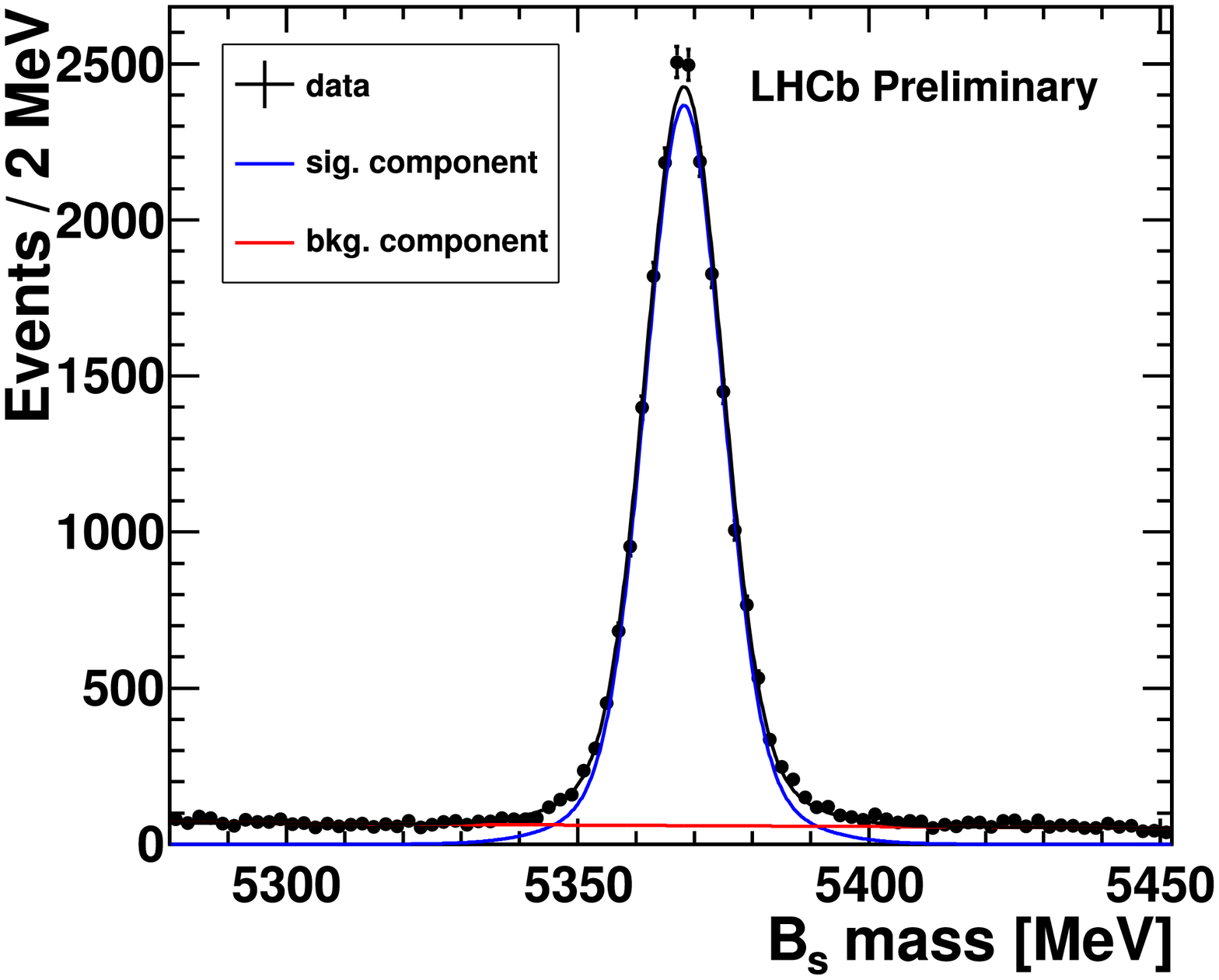}
\includegraphics[width=0.47\textwidth]{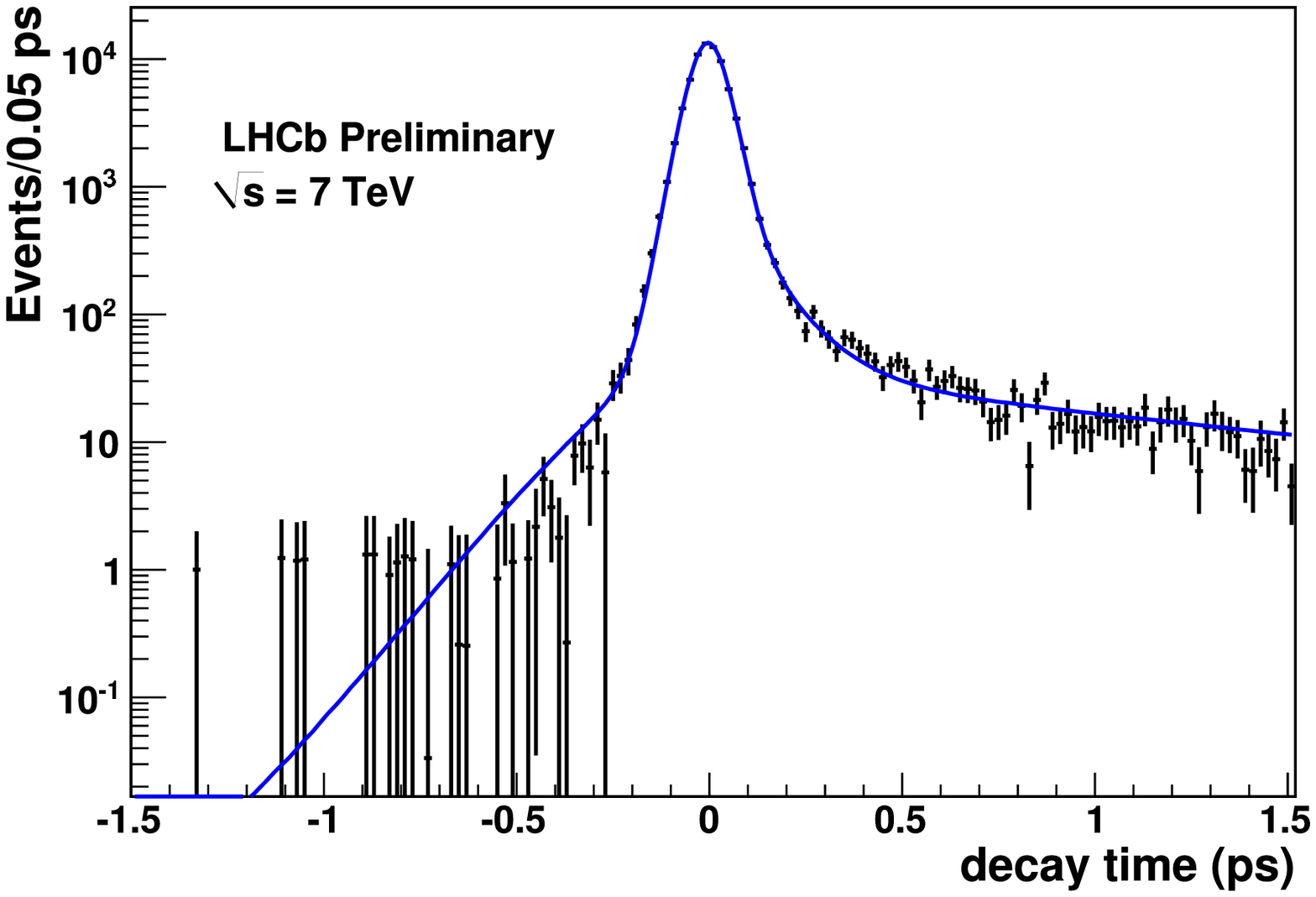}
\caption{
(Left) Reconstructed invariant mass distribution of the selected $\Bs \to J/\psi \phi$ candidates.
(Right) decay time distribution of $\Bs \to J/\psi \phi$ candidates with a true $J/\psi$ including the $t \sim 0$ region and superimposed 
fit result. Both are  from Ref.~\cite{LHCb-CONF-2012-002}.
}
\label{fig:lhcb_data}
\end{figure}

Recently, LHCb updated its $\Bs \to J/\psi \phi$ analysis result using   1 fb$^{-1}$ of $pp$ collision data collected during the 2011 LHC run
at a center of mass energy of $\sqrt{s} = 7 $ TeV~\cite{LHCb-CONF-2012-002}. A clean sample containing about 21,200 $\Bs \to J/\psi \phi$
signal events in a $K^+K^-$ mass window of 12 MeV around the $\phi$ mass peak  is selected using the  particle identification and kinematic information. 
The reconstructed invariant mass distribution of the
selected $\Bs \to J/\psi \phi$ candidates with decay time $t$ above 0.3 ps is shown in Fig.~\ref{fig:lhcb_data} (left),
where a $J/\psi$ mass constraint is applied in the vertex fit.
These events were triggered by requiring a relatively high transverse momentum  muon track from the $J/\psi$ decay to be
displaced from the $pp$ interaction point. The trigger efficiency depends on 
the decay time of the $\Bs$ mesons. The geometrical acceptance of the detector and the  kinematic requirements on the final state particles
also induce a dependence of the reconstruction efficiency as a function of the angular variables. Both efficiency effects are 
corrected for in the analysis. The background is dominated by combinatorial events and its 
decay time and angular model is constructed using  $\Bs$ mass sidebands.

The decay time resolution effect is modelled using a Gaussian model, which has a width $S_{\sigma t}\cdot \sigma_t$.
Here $\sigma_t$ is the event-by-event decay time resolution. The scale factor $S_{\sigma t}$  is estimated to be $1.45 \pm 0.06$  
from a fit to the  $t \sim 0$ region (Fig.~\ref{fig:lhcb_data} (right)) and allowed to vary within this uncertainty in the fit
for extraction of $\phi_s$.
This event-by-event resolution  model has a statistical power for measurement of mixing-induced $CP$ asymmetries in $\Bs$ decays
equivalent to that of a single Gaussian model with a constant width of 45 fs.

\begin{figure}[htb]
\centering
\includegraphics[width=0.42\textwidth,height=6.2cm]{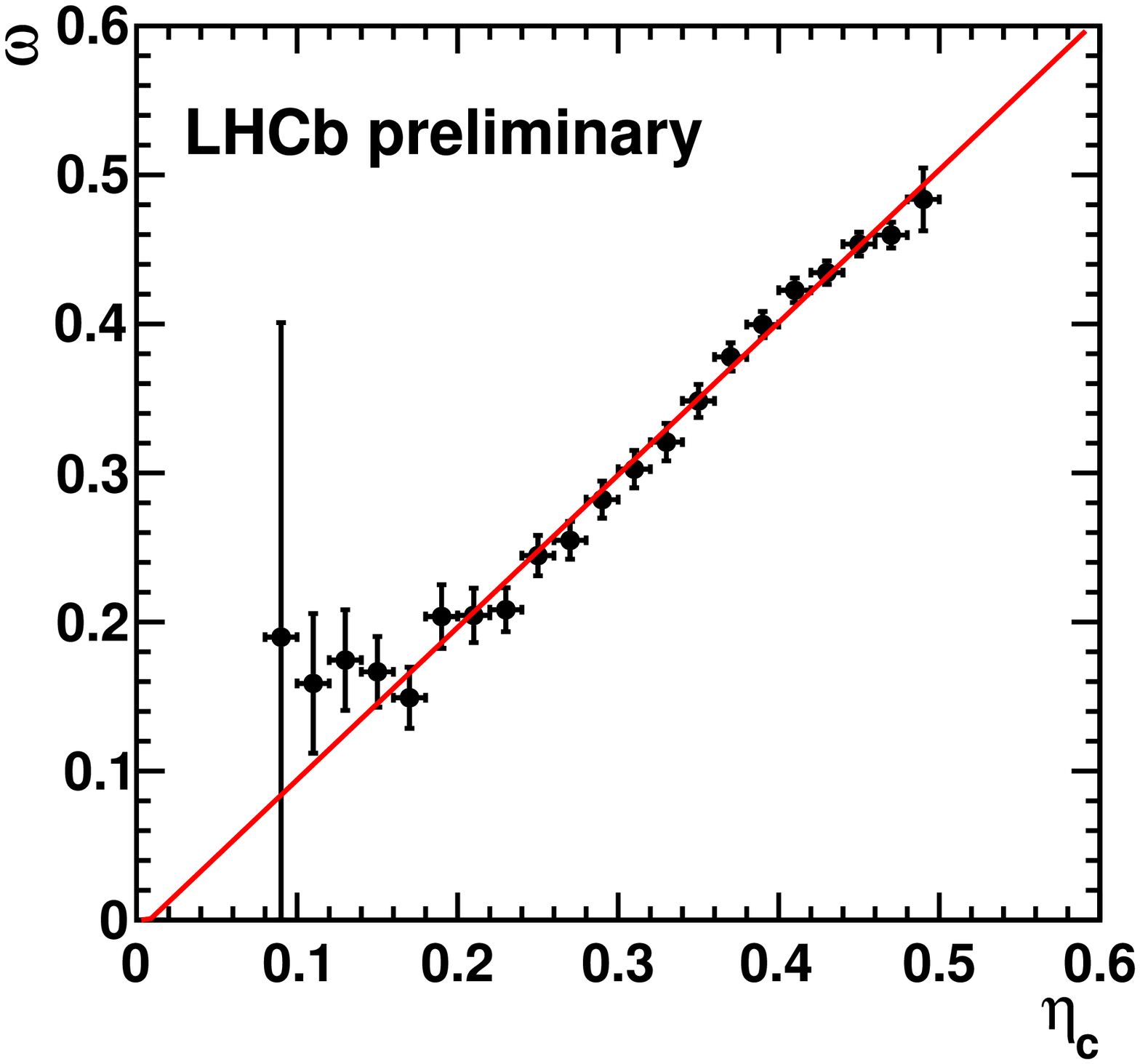}
\includegraphics[width=0.45\textwidth]{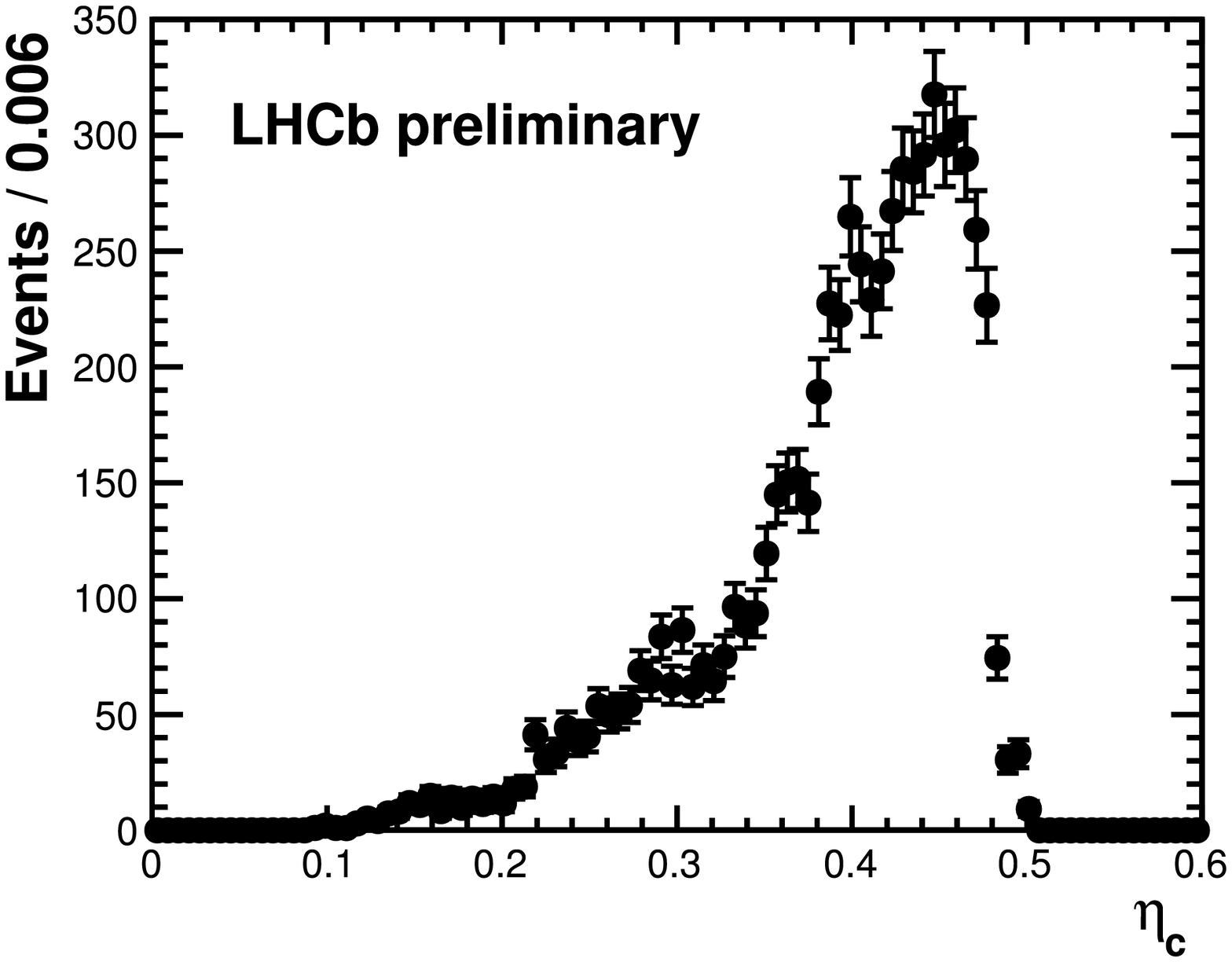}
\caption{
(Left) Measured OS wrong tag probability ($\omega$) as a function of the estimated OS wrong tag probability ($\eta_c$) for background 
subtracted $B^+ \to J/\psi K^+$  decays.
(Right) Distribution of calibrated OS wrong tag probability of tagged $\Bs \to J/\psi \phi$ signal candidates.
Both are  from Ref.~\cite{LHCb-CONF-2012-002}.
}
\label{fig:lhcb_tagging}
\end{figure}

In this analysis, the flavour of the $B$ (or ${\bar B}$) meson at production ($t=0$) is identified using the opposite side (OS) tagging method,
which exploits information about the other $b$-hadron from pair production of $b\bar b$ quarks, including  charges of the decay products of the other $b$-hadron.   
A wrong tag probability of the OS tagging decision is estimated for each $B$ candidate, and this estimated probability is calibrated in the control channel
$B^+ \to J/\psi K^+$, as shown in Fig.~\ref{fig:lhcb_tagging} (left). The distribution of calibrated OS wrong tag probability of the tagged $\Bs \to J/\psi \phi$ signal candidates
is shown in Fig.~\ref{fig:lhcb_tagging} (right), from which an average wrong tag probability of ${\bar \omega} = (36.81 \pm 0.18 (\rm stat) \pm 0.74(\rm syst))\%$   is obtained.
 The tagging efficiency is $\epsilon_{tag} = (32.99 \pm 0.33) \%$. 
The effective tagging efficiency for the $\Bs \to J/\psi \phi$  sample is estimated to be $\epsilon_{tag}(1-2{\bar \omega})^2 = (2.29 \pm 0.07 (\rm stat) \pm 0.26(\rm syst))\%$.

\begin{figure}[htb]
\centering
\includegraphics[width=0.4\textwidth]{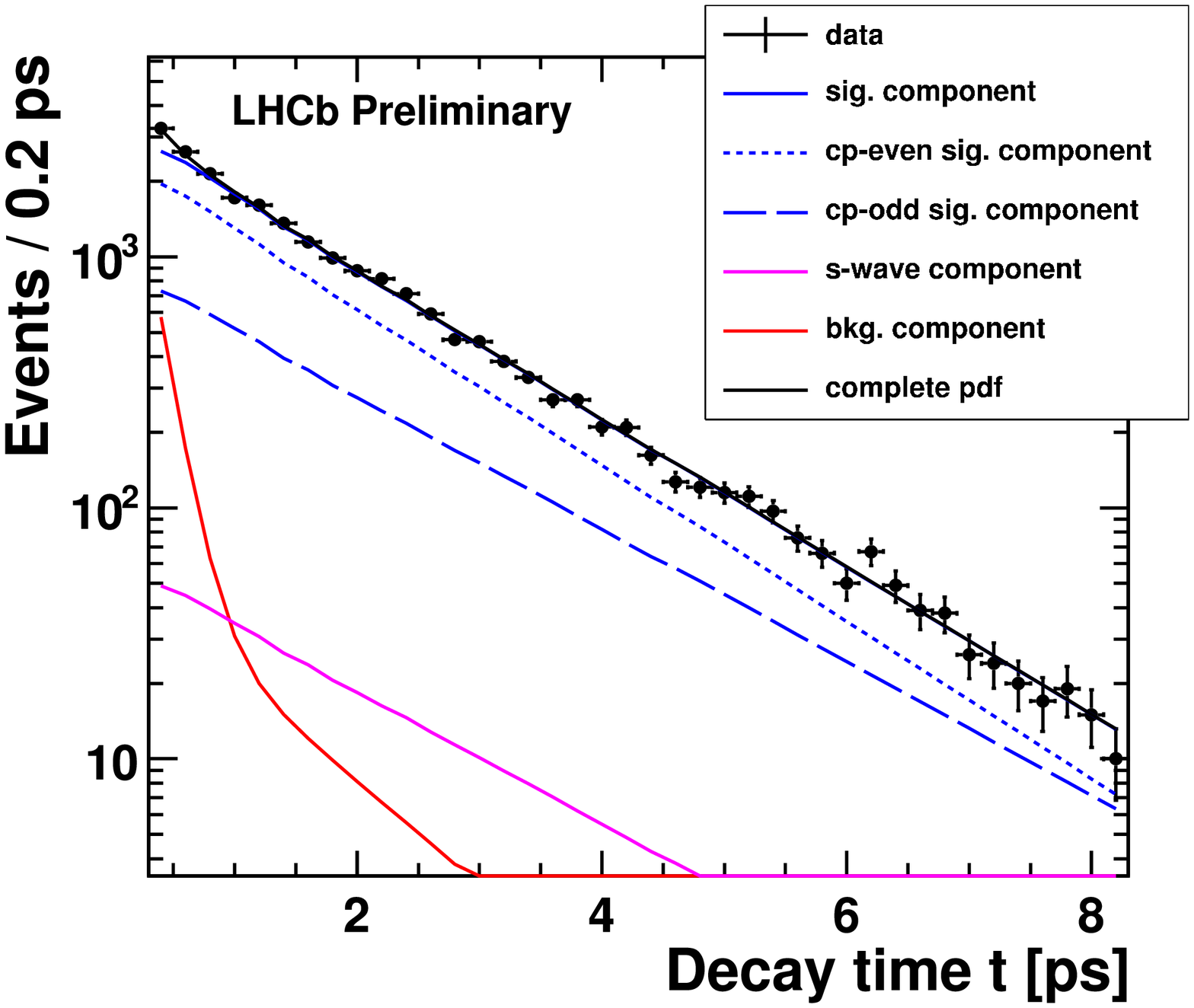}
\includegraphics[width=0.4\textwidth]{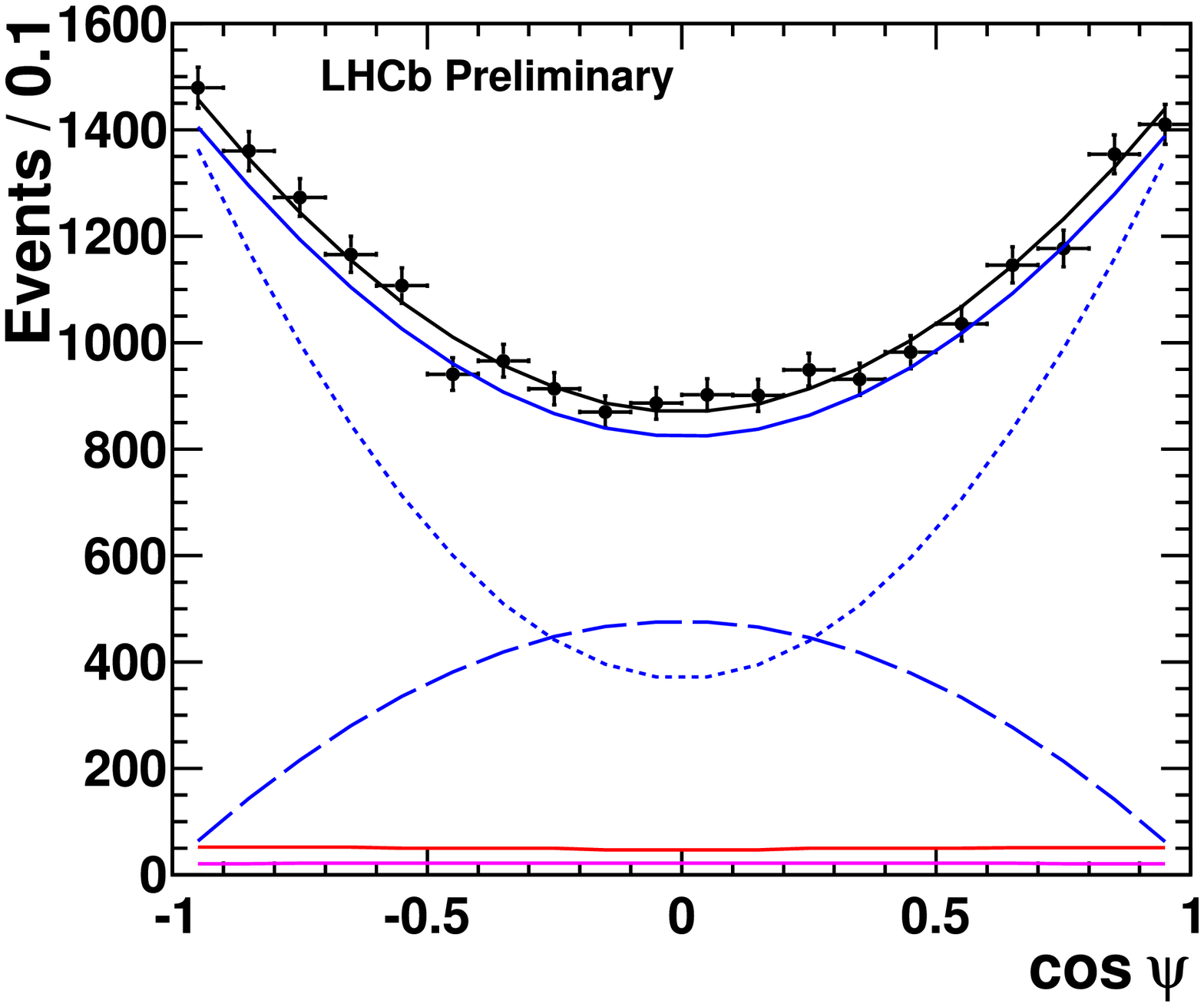}
\includegraphics[width=0.4\textwidth]{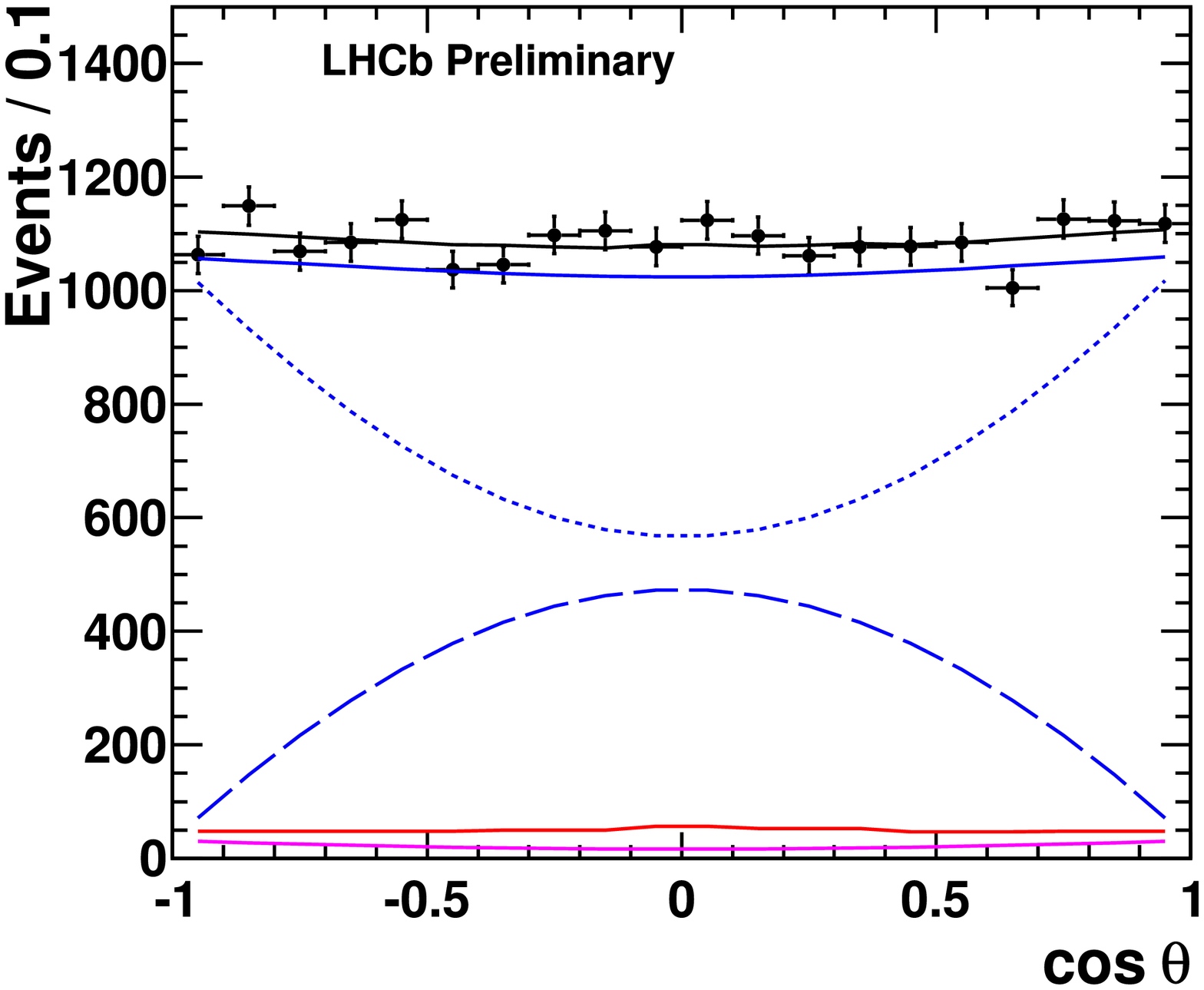}
\includegraphics[width=0.4\textwidth]{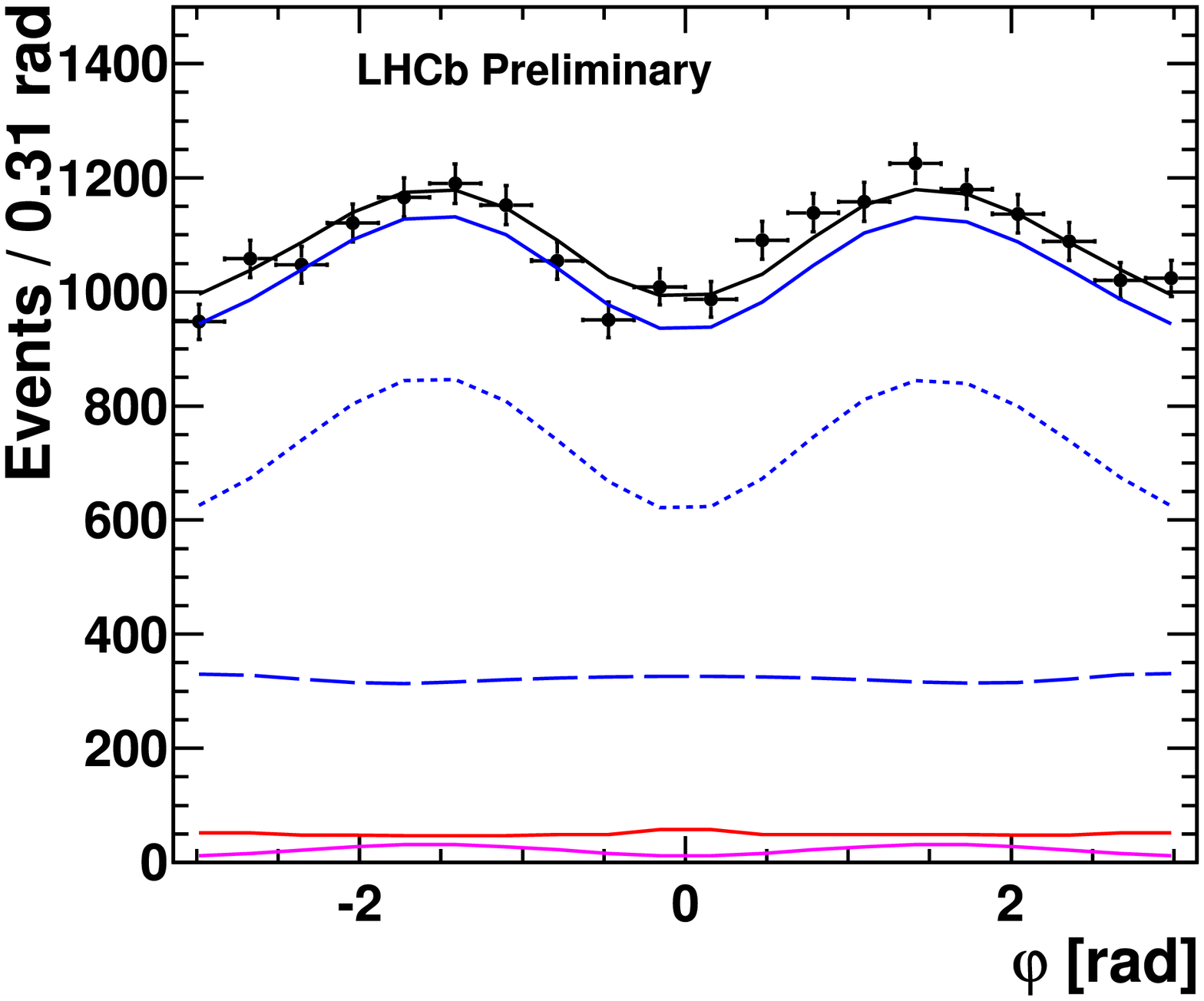}
\caption{
Data points and fit projections for the decay time and three angular variables for candidates in a $\pm 20$ MeV window around the  $\Bs$ mass peak~\cite{LHCb-CONF-2012-002}.
}
\label{fig:lhcb_fit}
\end{figure}

\begin{figure}[htb]
\centering
\includegraphics[width=0.51\textwidth]{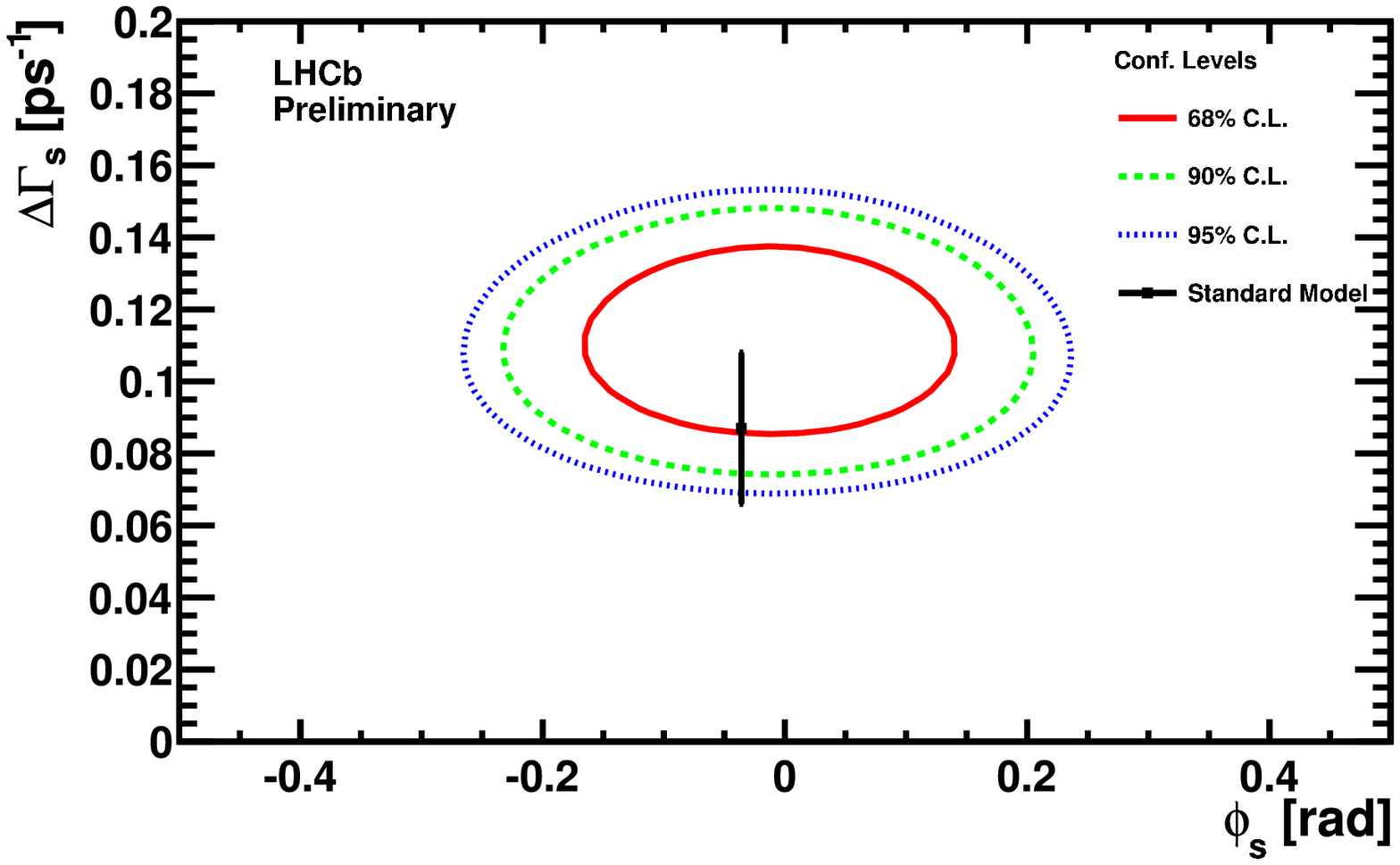}
\includegraphics[width=0.44\textwidth]{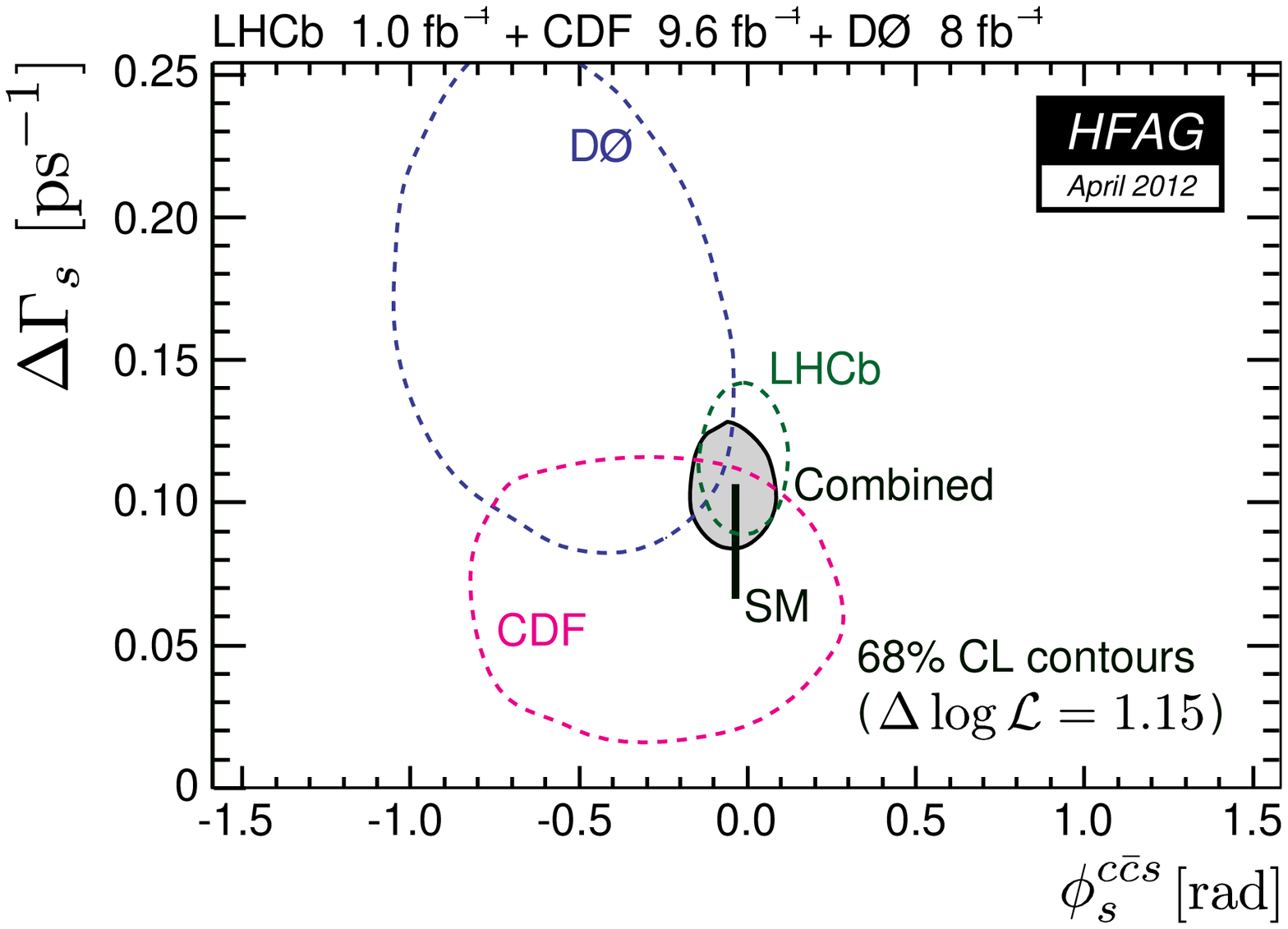}
\caption{
  (Left) LHCb measurement of $\phi_s$ and $\Delta\Gamma_s$ from $\Bs\to J/\psi\,\phi$ decays using 1.0 fb$^{-1}$~\cite{LHCb-CONF-2012-002}.
  (Right) HFAG 2012 combination of $\phi_s$ and $\Delta\Gamma_s$ results, where the 1 $\sigma$ confidence region is shown for each experiment and the combined result~\cite{HFAG}.
  \label{fig:bcpv:phis}
}
\label{fig:lhcb_result}
\end{figure}

The analysis uses the $\Bs$ oscillation frequency $\Delta m_s = 17.63 \pm 0.11$ $\text{ps}^{-1}$~\cite{LHCb-PAPER-2011-010} and allows it to float within its uncertainty.
The fit projections on the decay time $t$ and the three angular variables are shown in Fig.~\ref{fig:lhcb_fit}. 
The  numerical results of the major physics parameters are
\begin{equation}
  \begin{array}{ccllllllll}
    \phi_s &\;=\; & -0.001  &\pm & 0.101  & \text{(stat)} &\pm & 0.027 & \text{(syst)} & \text{rad},\\
    \Gamma_s &\;=\; & 0.6580 &\pm & 0.0054 & \text{(stat)} &\pm & 0.0066 & \text{(syst)} & \text{ps}^{-1},\rule{0pt}{5mm} \\
    \Delta\Gamma_s     &\;=\; & 0.116 &\pm & 0.018 & \text{(stat)} &\pm & 0.006 & \text{(syst)} & \text{ps}^{-1},\rule{0pt}{5mm} \\
    F_{\rm S}     &\;=\; & 0.022 &\pm & 0.012 & \text{(stat)} &\pm & 0.007 & \text{(syst)} &,\rule{0pt}{5mm} \\
  \end{array}
\end{equation}
where $F_{\rm S}$ is the fraction of S wave contribution in a window of 12 MeV around the $\phi$ mass.
As can be seen in Fig.~\ref{fig:bcpv:phis} (left), the measurement of  $\phi_s$ and $\Delta\Gamma_s$  is in good agreement with the SM predictions.
These results are still dominated  by statistical uncertainties. The important sources of systematic uncertainties include
the neglected direct $CP$ violation, insufficient modelling of angular acceptance and background effects. A refined LHCb analysis using the same data sample
is in progress. This  will include the same side kaon tagging information to  increase  effective tagging efficiency 
 and also benefit from the improved understanding of the  systematic uncertainties.

CDF  updated its $\Bs \to J/\psi \phi$ analysis using  9.6 fb$^{-1}$ of $p\bar p$ collision data collected 
at a center of mass energy of $\sqrt{s} = 1.96 $ TeV at the Tevatron~\cite{CDF-10778}.  
Approximately 11,000 signal decays are selected and analyzed. 
Due to the low decay time resolution, the CDF analysis has limited sensitivity to $\phi_s$.
The  CDF $\Delta\Gamma_s$ result has a precision comparable to that of the LHCb result:
\begin{equation}
 \Delta \Gamma_s = 0.068 \pm 0.026\,\, (\text{stat}) \pm 0.007\,\, (\text{syst})\,\, \text{ps}^{-1}.
\end{equation}

The latest HFAG~\cite{HFAG} average of  the  results from the LHCb~\cite{LHCb-CONF-2012-002}, CDF~\cite{CDF-10778} and D0~\cite{Abazov:2011ry} analyses
is shown in Fig.~\ref{fig:bcpv:phis} (right).
The LHCb result dominates the combination, which is in good agreement with the SM predictions.
(At the time of writing this article, the ATLAS experiment has also reported the result of
a time-dependent angular analysis of $\Bs\to J/\psi\,\phi$ decays without flavour tagging~\cite{2012fu}. We do not discuss this result here.)

\begin{figure}[!htb]
\centering
\includegraphics[width=0.48\textwidth]{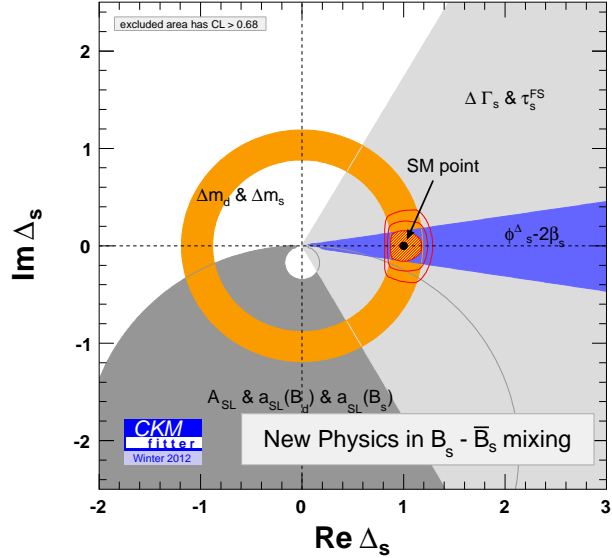}
\caption{
  Model-independent fit~\cite{Lenz:2012az} in the scenario that new physics affects $M^s_{12}$.
  The coloured areas represent regions with CL $< 68.3\,\%$ for the individual constraints.
  The red area shows the region with CL $< 68.3\,\%$ for the combined fit, with the two additional contours delimiting the regions with CL $< 95.45\,\%$ and CL $< 99.73\,\%$.
 \label{fig:bcpv:newphysics}
}
\end{figure}

In the context of model-independent analysis, the new physics contribution to $M^s_{12}$ can be parameterized using the complex number $\Delta_s$,
\begin{equation}
  \label{eq:NPcontribution:def}
  M_{12}^s  = M_{12}^{s,{\rm SM}} \, \Delta_s \,.
\end{equation}
The constraints on $\Delta_s$ provided by the current measurements of $\phi_s$, $\Delta m_s$, $\Delta m_d$, $\Delta\Gamma_s$ and semileptonic asymmetries,
are shown in Fig.~\ref{fig:bcpv:newphysics}~\cite{Lenz:2012az}. As can be seen, the major constraints on new physics in $M^s_{12}$ come from
the measurements of $\phi_s$ and $\Delta m_{s/d}$. No significant new physics contribution is identified and the picture of $\Bs$ mixing is SM-like.
However, up to about 30$\%$ new physics contribution in $M^s_{12}$  is still allowed at 3$\sigma$ confidence level,
and probing new physics in $\Bs$ mixing at this level requires to improve substantially the measurement precision of $\phi_s$. 
The LHCb experiment is expected to collect 5 fb$^{-1}$ of data before 2018 and  50 fb$^{-1}$  after its upgrade.
This will enable LHCb to push down the uncertainty of $\phi_s$ to $\sim 0.025$ rad around 2018, and eventually achieve a precision of
  $\sim 0.008$ rad after the upgrade~\cite{lhcb-upgradeLOI, lhcb-upgradeFTDR}.

\section{Conclusions}

In summary, the study of $CP$ violation in $\Bs \to J/\psi \phi$  offers a great opportunity to search for physics beyond the SM that enters the $\Bs$ mixing process.
Significant progress has been made in the measurement of $\phi_s$ in $\Bs \to J/\psi \phi$, particularly by the LHCb experiment, that allows to put stringent constraints
on new physics contribution in $\Bs$ mixing. The LHCb experiment aims to greatly improve the $\phi_s$ measurement precision for probing sub-leading level new physics contribution 
in $\Bs$ mixing. 


\bigskip 
\newpage

\begin{thebibliography}{19}
\expandafter\ifx\csname natexlab\endcsname\relax\def\natexlab#1{#1}\fi
\expandafter\ifx\csname bibnamefont\endcsname\relax
  \def\bibnamefont#1{#1}\fi
\expandafter\ifx\csname bibfnamefont\endcsname\relax
  \def\bibfnamefont#1{#1}\fi
\expandafter\ifx\csname citenamefont\endcsname\relax
  \def\citenamefont#1{#1}\fi
\expandafter\ifx\csname url\endcsname\relax
  \def\url#1{\texttt{#1}}\fi
\expandafter\ifx\csname urlprefix\endcsname\relax\def\urlprefix{URL }\fi
\providecommand{\bibinfo}[2]{#2}
\providecommand{\eprint}[2][]{\url{#2}}

\bibitem[{\citenamefont{Bediaga et~al.}(2012)}]{Bediaga:2012py}
\bibinfo{author}{\bibfnamefont{I.}~\bibnamefont{Bediaga}} \bibnamefont{et~al.}
  (\bibinfo{collaboration}{LHCb collaboration}) (\bibinfo{year}{2012}),
  \eprint{1208.3355}.

\bibitem[{\citenamefont{Lenz and Nierste}(2007)}]{Lenz:2006hd}
\bibinfo{author}{\bibfnamefont{A.}~\bibnamefont{Lenz}} \bibnamefont{and}
  \bibinfo{author}{\bibfnamefont{U.}~\bibnamefont{Nierste}},
  \bibinfo{journal}{JHEP} \textbf{\bibinfo{volume}{06}}, \bibinfo{pages}{072}
  (\bibinfo{year}{2007}), \eprint{hep-ph/0612167}.

\bibitem[{\citenamefont{Charles et~al.}(2011)\citenamefont{Charles, Deschamps,
  Descotes-Genon, Itoh, Lacker et~al.}}]{Charles:2011va}
\bibinfo{author}{\bibfnamefont{J.}~\bibnamefont{Charles}},
  \bibinfo{author}{\bibfnamefont{O.}~\bibnamefont{Deschamps}},
  \bibinfo{author}{\bibfnamefont{S.}~\bibnamefont{Descotes-Genon}},
  \bibinfo{author}{\bibfnamefont{R.}~\bibnamefont{Itoh}},
  \bibinfo{author}{\bibfnamefont{H.}~\bibnamefont{Lacker}},
  \bibnamefont{et~al.}, \bibinfo{journal}{Phys.Rev.}
  \textbf{\bibinfo{volume}{D84}}, \bibinfo{pages}{033005}
  (\bibinfo{year}{2011}), \eprint{1106.4041}.

\bibitem[{\citenamefont{Dighe et~al.}(1999)\citenamefont{Dighe, Dunietz, and
  Fleischer}}]{Dighe:1998vk}
\bibinfo{author}{\bibfnamefont{A.~S.} \bibnamefont{Dighe}},
  \bibinfo{author}{\bibfnamefont{I.}~\bibnamefont{Dunietz}}, \bibnamefont{and}
  \bibinfo{author}{\bibfnamefont{R.}~\bibnamefont{Fleischer}},
  \bibinfo{journal}{Eur.Phys.J.} \textbf{\bibinfo{volume}{C6}},
  \bibinfo{pages}{647} (\bibinfo{year}{1999}), \eprint{hep-ph/9804253}.

\bibitem[{\citenamefont{Dunietz et~al.}(2001)\citenamefont{Dunietz, Fleischer,
  and Nierste}}]{Dunietz:2000cr}
\bibinfo{author}{\bibfnamefont{I.}~\bibnamefont{Dunietz}},
  \bibinfo{author}{\bibfnamefont{R.}~\bibnamefont{Fleischer}},
  \bibnamefont{and} \bibinfo{author}{\bibfnamefont{U.}~\bibnamefont{Nierste}},
  \bibinfo{journal}{Phys.Rev.} \textbf{\bibinfo{volume}{D63}},
  \bibinfo{pages}{114015} (\bibinfo{year}{2001}), \eprint{hep-ph/0012219}.

\bibitem[{\citenamefont{Xie et~al.}(2009)\citenamefont{Xie, Clarke, Cowan, and
  Muheim}}]{Xie:2009fs}
\bibinfo{author}{\bibfnamefont{Y.}~\bibnamefont{Xie}},
  \bibinfo{author}{\bibfnamefont{P.}~\bibnamefont{Clarke}},
  \bibinfo{author}{\bibfnamefont{G.}~\bibnamefont{Cowan}}, \bibnamefont{and}
  \bibinfo{author}{\bibfnamefont{F.}~\bibnamefont{Muheim}},
  \bibinfo{journal}{JHEP} \textbf{\bibinfo{volume}{0909}}, \bibinfo{pages}{074}
  (\bibinfo{year}{2009}), \eprint{0908.3627}.

\bibitem[{D0-()}]{D0-5928}
\bibinfo{note}{CDF and D0 collaborations, D0 Note 5928-CONF}.

\bibitem[{\citenamefont{Aaltonen et~al.}(2012)}]{CDF:2011af}
\bibinfo{author}{\bibfnamefont{T.}~\bibnamefont{Aaltonen}} \bibnamefont{et~al.}
  (\bibinfo{collaboration}{CDF collaboration}), \bibinfo{journal}{Phys.Rev.}
  \textbf{\bibinfo{volume}{D85}}, \bibinfo{pages}{072002}
  (\bibinfo{year}{2012}), \eprint{1112.1726}.

\bibitem[{\citenamefont{Abazov et~al.}(2012)}]{Abazov:2011ry}
\bibinfo{author}{\bibfnamefont{V.~M.} \bibnamefont{Abazov}}
  \bibnamefont{et~al.} (\bibinfo{collaboration}{D0 collaboration}),
  \bibinfo{journal}{Phys.Rev.} \textbf{\bibinfo{volume}{D85}},
  \bibinfo{pages}{032006} (\bibinfo{year}{2012}), \eprint{1109.3166}.

\bibitem[{\citenamefont{Aaij et~al.}(2012{\natexlab{a}})}]{LHCb-PAPER-2011-021}
\bibinfo{author}{\bibfnamefont{R.}~\bibnamefont{Aaij}} \bibnamefont{et~al.}
  (\bibinfo{collaboration}{LHCb collaboration}),
  \bibinfo{journal}{Phys.Rev.Lett.} \textbf{\bibinfo{volume}{108}},
  \bibinfo{pages}{101803} (\bibinfo{year}{2012}{\natexlab{a}}),
  \eprint{1112.3183}.

\bibitem[{\citenamefont{Aaij et~al.}(2012{\natexlab{b}})}]{LHCb-PAPER-2011-028}
\bibinfo{author}{\bibfnamefont{R.}~\bibnamefont{Aaij}} \bibnamefont{et~al.}
  (\bibinfo{collaboration}{LHCb collaboration}),
  \bibinfo{journal}{Phys.Rev.Lett.} \textbf{\bibinfo{volume}{108}},
  \bibinfo{pages}{241801} (\bibinfo{year}{2012}{\natexlab{b}}),
  \eprint{1202.4717}.

\bibitem[{LHC()}]{LHCb-CONF-2012-002}
\bibinfo{note}{LHCb collaboration, LHCb-CONF-2012-002}.

\bibitem[{\citenamefont{Aaij et~al.}(2012{\natexlab{c}})}]{LHCb-PAPER-2011-010}
\bibinfo{author}{\bibfnamefont{R.}~\bibnamefont{Aaij}} \bibnamefont{et~al.}
  (\bibinfo{collaboration}{LHCb collaboration}), \bibinfo{journal}{Phys. Lett.}
  \textbf{\bibinfo{volume}{B709}}, \bibinfo{pages}{177}
  (\bibinfo{year}{2012}{\natexlab{c}}), \eprint{1112.4311}.

\bibitem[{CDF()}]{CDF-10778}
\bibinfo{note}{CDF collaboration, CDF note 10778.}

\bibitem[{\citenamefont{Amhis et~al.}(2012)}]{HFAG}
\bibinfo{author}{\bibfnamefont{Y.}~\bibnamefont{Amhis}} \bibnamefont{et~al.}
  (\bibinfo{collaboration}{Heavy Flavor Averaging Group})
  (\bibinfo{year}{2012}), \bibinfo{note}{{updated results and plots available
  at: \href{http://www.slac.stanford.edu/xorg/hfag/}{{\tt
  http://www.slac.stanford.edu/xorg/hfag/}}}}, \eprint{1207.1158}.

\bibitem[{\citenamefont{Aad et~al.}(2012)}]{2012fu}
\bibinfo{author}{\bibfnamefont{G.}~\bibnamefont{Aad}} \bibnamefont{et~al.}
  (\bibinfo{collaboration}{ATLAS Collaboration}) (\bibinfo{year}{2012}),
  \eprint{1208.0572}.

\bibitem[{\citenamefont{Lenz et~al.}(2012)\citenamefont{Lenz, Nierste, Charles,
  Descotes-Genon, Lacker et~al.}}]{Lenz:2012az}
\bibinfo{author}{\bibfnamefont{A.}~\bibnamefont{Lenz}},
  \bibinfo{author}{\bibfnamefont{U.}~\bibnamefont{Nierste}},
  \bibinfo{author}{\bibfnamefont{J.}~\bibnamefont{Charles}},
  \bibinfo{author}{\bibfnamefont{S.}~\bibnamefont{Descotes-Genon}},
  \bibinfo{author}{\bibfnamefont{H.}~\bibnamefont{Lacker}},
  \bibnamefont{et~al.} (\bibinfo{year}{2012}), \eprint{1203.0238}.

\bibitem[{lhc({\natexlab{a}})}]{lhcb-upgradeLOI}
\bibinfo{note}{LHCb collaboration, CERN-LHCC-2011-001.}

\bibitem[{lhc({\natexlab{b}})}]{lhcb-upgradeFTDR}
\bibinfo{note}{LHCb collaboration, CERN-LHCC-2012-027.}

\end{thebibliography}

\end{document}